# Master equation for retrodiction of quantum communication signals

Stephen M. Barnett[1], David T. Pegg[2], John Jeffers[1] and Ottavia Jedrkiewicz[3]

[1]Department of Physics and Applied Physics, University of Strathclyde,

Glasgow G4 0NG, U.K.

[2]School of Science, Griffith University, Nathan, Brisbane 4111, Australia.

[3]Department of Electronic Systems Engineering, University of Essex,

Wivenhoe Park, Colchester CO4 3SQ, U.K.

We derive the master equation that governs the evolution of the measured state backwards in time in an open system. This allows us to determine probabilities for a given set of preparation events from the results of subsequent measurements, which has particular relevance to quantum communication.

The retrodictive formalism of quantum mechanics has been studied for some time but until recently it has been mainly an interesting philosophical concept associated with the problem of time asymmetry in quantum mechanics [1]. With the recent rapid development and interest in quantum communications [2], including quantum cryptography [3], however, retrodictive quantum mechanics will become increasingly important [4]. The essential communication problem is to determine the message sent from the signal received [5]. The basic quantum communication problem is as follows. A quantum system is prepared in some state by Alice and sent to Bob. Bob must retrodict from the output of his measurement or detection device the signal state selected by Alice. For this he needs to be able to calculate, on the basis of a single detection event, the probability that Alice prepared any particular state from a set of known possible states. This problem can be solved using the predictive formalism of quantum mechanics combined with Bayes' theorem [6], which relates predictive and retrodictive conditional probabilities. Thus the probability that a particular preparation event $i$ occurred given that a measurement provides the event $j$ is

$$P(i|j) = \frac{P(j|i)P(i)}{P(j)} = \frac{P(j|i)P(i)}{\sum_i P(j|i)P(i)}. \tag{1}$$

The retrodictive formalism, however, provides a more direct and natural approach. For closed systems, in which the evolution of the quantum state is unitary between preparation and measurement, the intrinsic time symmetry simplifies the problem significantly. Open systems, on the other hand, in which the quantum system of interest interacts with a large environment into which information is irretrievably lost, provide a more realistic model of practical situations. For these systems the simplifying assumption of time symmetry is no longer applicable. We have shown elsewhere [7, 8] how one can exploit Bayes' theorem in conjunction with the solution of a predictive master equation to address the problem of retrodiction in open systems. In this paper we derive a general *retrodictive* master equation that can be applied directly to the basic quantum communication problem described above.

We can represent the quantum measuring, or detecting, device mathematically by means of a probability operator measure (POM) with elements $\hat{\Pi}_j$ which sum to the unit operator [9]. The expectation values of these elements are the probabilities for the corresponding possible measurement outcomes $j$. In the predictive picture of quantum mechanics, for a closed system prepared by Alice at time $t_p$ in state $\hat{\rho}_i^{\text{pred}}(t_p)$ the predictive conditional probability of Bob obtaining the measurement outcome $j$ at the later measurement time $t_m$ is

$$P(j|i) = \text{Tr}\left[\hat{U}(t_m,t_p)\hat{\rho}_i^{\text{pred}}(t_p)\hat{U}^\dagger(t_m,t_p)\hat{\Pi}_j\right] = \text{Tr}\left[\hat{\rho}_i^{\text{pred}}(t_m)\hat{\Pi}_j\right], \tag{2}$$

where $\hat{U}(t_m,t_p)$ is the evolution operator from $t_p$ to $t_m$. We note here that if we write the evolution operator as the product of two separate operators, $\hat{U}(t_m,t_p) = \hat{U}(t_m,t)\hat{U}(t,t_p)$ for any time $t$ between preparation and measurement, and use the cyclic property of the trace the predictive conditional probability can be written as

$$P(j|i) = \text{Tr}\left[\hat{\rho}_i^{\text{pred}}(t)\hat{\Pi}_j(t)\right]. \tag{3}$$

Here $\hat{\Pi}_j(t)$ is the measurement POM element evolved *backwards* in time from $t_m$ to the intermediate time $t$. The conditional probability is independent of this intermediate time.

In this paper we are interested in the more difficult, but more practical, case of an open system. Here the simple time symmetry inherent in the unitary evolution no longer applies. This is because in both the predictive and retrodictive formalisms the initial state of the environment is known and the measurement provides no information about the final environment state. We can treat the problem as that of a closed system comprising the environment *E* and the system of interest *S* as subsystems. We write the POM element for outcome *j* of a measurement on *S* as $\hat{\Pi}_{j,S}$. The environment POM elements must sum to $\hat{1}_E$, the unit operator on the state space of *E*. As more than one possible outcome of a measurement on *E* would provide some information about the environment, the environment POM element must be $\hat{1}_E$. The POM elements $\hat{\Pi}_j$ for the closed system, of system plus environment, are thus $\hat{1}_E \otimes \hat{\Pi}_{j,S}$. The predictive conditional probability for Bob to obtain the outcome *j* if Alice prepares the state $\hat{\rho}_i^{\text{pred}}(t_p)$ is then

$$P(j|i) = \text{Tr}_{ES}\left[\hat{U}(t_m,t_p)\hat{\rho}_E^{\text{pred}}(t_p) \otimes \hat{\rho}_{i,S}^{\text{pred}}(t_p)\hat{U}^\dagger(t_m,t_p)\hat{1}_E \otimes \hat{\Pi}_{j,S}\right], \tag{4}$$

where $\hat{U}(t_m,t_p)$ is the evolution operator for the total system and environment combined. This equation can also be written in terms of a general intermediate time *t* as

$$P(j|i) = \text{Tr}_{ES}\left[\hat{U}(t,t_p)\hat{\rho}_E^{\text{pred}}(t_p) \otimes \hat{\rho}_{i,S}^{\text{pred}}(t_p)\hat{U}^\dagger(t,t_p)\hat{U}^\dagger(t_m,t)\hat{1}_E \otimes \hat{\Pi}_{j,S}\hat{U}(t_m,t)\right]. \tag{5}$$

The standard predictive master equation approximation is that the environment has a large number of degrees of freedom and is little changed by the coupling to *S*, that is [10]

$$\hat{U}(t,t_p)\hat{\rho}_E^{\text{pred}}(t_p) \otimes \hat{\rho}_{i,S}^{\text{pred}}(t_p)\hat{U}^\dagger(t,t_p) \approx \hat{\rho}_E^{\text{pred}}(t_p) \otimes \hat{\rho}_{i,S}^{\text{pred}}(t) \tag{6}$$

where

$$\hat{\rho}_{i,S}^{\text{pred}}(t) = \text{Tr}_E[\hat{U}(t,t_p)\hat{\rho}_E^{\text{pred}}(t_p) \otimes \hat{\rho}_{i,S}^{\text{pred}}(t_p)\hat{U}^\dagger(t,t_p)] \tag{7}$$

is the reduced predictive density operator for the system *S* [10]. We use this to write Eq. (5) as

$$P(j|i) = \text{Tr}_S\left[\hat{\rho}_{i,S}^{\text{pred}}(t)\hat{\Pi}_{j,S}(t)\right] \tag{8}$$

where

$$\hat{\Pi}_{j,S}(t) = \text{Tr}_E[\hat{\rho}_E^{\text{pred}}(t_p) \otimes \hat{U}^\dagger(t_m,t)\hat{\Pi}_{j,S}\hat{U}(t_m,t)] \ . \tag{9}$$

As is the case for closed systems, the probability $P(j|i)$ in Eq.(5) is independent of *t*, provided we choose it somewhere between $t_p$ and $t_m$, so $\dot{P}(j|i)$, its time derivative with respect to *t*, is zero. Thus from (8)

$$\text{Tr}_S\left[\hat{\rho}_{i,S}^{\text{pred}}(t)\dot{\hat{\Pi}}_{j,S}(t)\right] = -\text{Tr}_S\left[\dot{\hat{\rho}}_{i,S}^{\text{pred}}(t)\hat{\Pi}_{j,S}(t)\right] \ . \tag{10}$$

We now consider the general predictive Markovian master equation for $\hat{\rho}_{i,S}^{\text{pred}}(t)$ in the standard Lindblad form [11]

$$\dot{\hat{\rho}}_{i,S}^{\text{pred}}(t) = -i\hbar^{-1}[\hat{H}_S, \hat{\rho}_{i,S}^{\text{pred}}(t)] + \sum_q \left[ 2\hat{A}_q \hat{\rho}_{i,S}^{\text{pred}}(t)\hat{A}_q^\dagger - \hat{A}_q^\dagger \hat{A}_q \hat{\rho}_{i,S}^{\text{pred}}(t) - \hat{\rho}_{i,S}^{\text{pred}}(t)\hat{A}_q^\dagger \hat{A}_q \right] \quad (11)$$

where $\hat{H}_S$ is the Hamiltonian for the system without the environment, and $\hat{A}_q$ is a system operator. We find, upon using the cyclic property of the trace,

$$\text{Tr}_S\left[ \hat{\rho}_{i,S}^{\text{pred}}(t) \dot{\hat{\Pi}}_{j,S}(t) \right]$$

$$= \text{Tr}_S\left[ \hat{\rho}_{i,S}^{\text{pred}}(t) \left\{ -i\hbar^{-1}[\hat{H}_S, \hat{\Pi}_j(t)] - \sum_q \left[ 2\hat{A}_q^\dagger \hat{\Pi}_{j,S}(t)\hat{A}_q - \hat{\Pi}_{j,S}(t)\hat{A}_q^\dagger \hat{A}_q - \hat{A}_q^\dagger \hat{A}_q \hat{\Pi}_{j,S}(t) \right] \right\} \right]. \quad (12)$$

This is true for all $\hat{\rho}_{i,S}^{\text{pred}}(t)$ so the evolution equation for the POM element is

$$\dot{\hat{\Pi}}_{j,S}(t) = -i\hbar^{-1}[\hat{H}_S, \hat{\Pi}_{j,S}(t)] - \sum_q \left[ 2\hat{A}_q^\dagger \hat{\Pi}_{j,S}(t)\hat{A}_q - \hat{\Pi}_{j,S}(t)\hat{A}_q^\dagger \hat{A}_q - \hat{A}_q^\dagger \hat{A}_q \hat{\Pi}_{j,S}(t) \right]. \quad (13)$$

We note that the evolution of $\hat{\dot{\Pi}}_{j,S}(t)$ is always backwards in time, that is Eq. (13) only holds for times $t \leq t_m$. The derivative of $\hat{\dot{\Pi}}_{j,S}(t)$ with respect to the premeasurement time, defined as $\tau = t_m - t$, is the negative of Eq. (13).

The retrodictive formalism is most useful for calculating retrodictive conditional probabilities $P(i|j)$ rather than predictive probabilities $P(j|i)$. These can be related using Bayes' theorem, so from Eqs. (1) and (8)

$$P(i|j) = \frac{\text{Tr}_S\left[\hat{\rho}_{i,S}^{\text{pred}}(t)\hat{\Pi}_{j,S}(t)\right]P(i)}{\sum_i \text{Tr}_S\left[\hat{\rho}_{i,S}^{\text{pred}}(t)\hat{\Pi}_{j,S}(t)\right]P(i)} = \frac{\text{Tr}_S\left[\hat{\rho}_{j,S}^{\text{retr}}(t)\hat{\Lambda}_{i,S}(t)\right]}{\sum_i \text{Tr}_S\left[\hat{\rho}_{j,S}^{\text{retr}}(t)\hat{\Lambda}_{i,S}(t)\right]}, \quad (14)$$

where

$$\rho_{j,S}^{\text{retr}}(t) = \frac{\hat{\Pi}_{j,S}(t)}{\text{Tr}_S\left[\hat{\Pi}_{j,S}(t)\right]} \quad (15)$$

is the retrodictive density operator describing the system at time *t*, which has been evolved backwards in time from the measurement time, and

$$\hat{\Lambda}_{i,S}(t) = P(i)\hat{\rho}_{i,S}^{\text{pred}}(t) \quad (16)$$

is the preparation device operator $\hat{\Lambda}_i(t_p)$ evolved forwards from the preparation time. Each preparation device operator is the product of the density operator representing the associated output state in the predictive formalism and the *a priori* probability of it occurring. The sum of the operators $\hat{\Lambda}_i(t_p)$ is thus the *a priori* density operator that Bob would ascribe to the system in the predictive formalism immediately after preparation in the absence of any knowledge of the selection made by Alice or the result of his measurement [7].

We can now substitute Eq. (13) into the time derivative of Eq. (15) to obtain eventually

$$\dot{\hat{\rho}}_{j,S}^{retr}(t) = -i\hbar^{-1}[\hat{H}_S, \hat{\rho}_{j,S}^{retr}(t)] - \sum_q \left[ 2\hat{A}_q^\dagger \hat{\rho}_{j,S}^{retr}(t)\hat{A}_q - \hat{\rho}_{j,S}^{retr}(t)\hat{A}_q^\dagger \hat{A}_q - \hat{A}_q^\dagger \hat{A}_q \hat{\rho}_{j,S}^{retr}(t) \right]$$

$$-2\hat{\rho}_{j,S}^{retr}(t)\text{Tr}_S\{\hat{\rho}_{j,S}^{retr}(t)\sum_q [\hat{A}_q^\dagger, \hat{A}_q]\} \qquad (17)$$

as the desired master equation.

It is easy to show that Eq. (17) preserves the trace of $\hat{\rho}_{j,S}^{retr}(t)$. It is also possible to show that $\hat{\rho}_{j,S}^{retr}(t)$ is a non-negative definite operator for any time *t* between $t_p$ and $t_m$, provided $\hat{\Pi}_{j,S}(t)$ is non-negative definite, as it must be in order to be a POM element. A detailed proof of this will be given elsewhere. This proves that the master equation is a physical one. One consequence of this is that the non-negativity of $\hat{\Pi}_{j,S}(t)$, combined with the fact that the sum of $\hat{\Pi}_{j,S}(t)$ is the unit operator [12], ensures that the $\hat{\Pi}_{j,S}(t)$ are also the elements of a POM. Thus a projection of a state of *S* onto $\hat{\Pi}_{j,S}(t)$ formally represents a measurement of *S*. This allows the interpretation of the projections in Eqs. (8) and (3) as a measurement, and thus a collapse, taking place at any time *t* between $t_p$ and $t_m$. The predictive prepared state evolves continuously until the collapse time and the POM element $\hat{\Pi}_{j,S}(t)$ evolves back continuously until the collapse time. The physical interpretation of $\dot{P}(j|i) = 0$ is, therefore, that measurable probabilities are independent of when we choose the collapse time, underlining the somewhat arbitrary nature of the concept.

As a specific simple example of a retrodictive master equation, we look at the case of Alice sending to Bob a decohering qubit in the form of a two-level atom undergoing spontaneous emission into the environmental vacuum with decay from state $|e\rangle$ to $|g\rangle$. For convenience we write $\hat{\rho}_{j,S}^{\text{retr}}(t) = \hat{\rho}$ and work in terms of the pre-measurement time $\tau$, defined earlier as $t_m - t$. We find the retrodictive master equation

$$d\hat{\rho}/d\tau = \gamma \left[ |e\rangle\langle e|\rho_{gg} - \hat{\rho}|e\rangle\langle e|/2 - |e\rangle\langle e|\hat{\rho}/2 + \hat{\rho}(\rho_{ee} - \rho_{gg}) \right] \qquad (18)$$

where $\gamma$ is the decay constant. Suppose Bob detects the atom to be in the superposition state $|+\rangle = (|e\rangle + |g\rangle)/\sqrt{2}$. Here we solve (18) with $\hat{\rho}$ at $\tau = 0$ equal to the POM element $\hat{\Pi}_{j,S} = |+\rangle\langle+|$ and obtain

$$\hat{\rho}(\tau) = [1 + (|e\rangle\langle g| + |g\rangle\langle e|)\exp(-\gamma\tau/2)]/2 . \tag{19}$$

If we know that Alice prepares the atom in state $|+\rangle$ and its orthogonal state $|-\rangle$ with equal probabilities, then the preparation device operator $\hat{\Lambda}_{+,S}$ corresponding to $|+\rangle$ is just $|+\rangle\langle+|/2$ and $\hat{\Lambda}_{-,S} = |-\rangle\langle-|/2$. Substitution into (4) gives the probability that Alice prepared state $|+\rangle$ at time $t_p$ as $\{1 + \exp[-\gamma(t_m - t_p)/2]\}/2$. For short pre-measurement times this is unity and for very long times it tends to 1/2, the *a priori* value in the absence of any measurement information. These results are precisely in accord with that what we have found recently by a more direct appeal to Bayes' theorem [8]. We should note here that, while solving the master equation (17) is straightforward in this case, for more complicated examples it may be easier to solve the corresponding retrodictive equation (13) for the POM element, and then normalise the solution by means of Eq. (15) to find the retrodictive density matrix. Also, in general Eq. (17) will be non-linear, while Eq.(13) will be linear, so Eq.(13) might be regarded as the more useful formulation of retrodictive evolution.

In conclusion, we have examined the basic problem of quantum communication for an open system, that is to retrodict the quantum state sent by Alice on the basis of a single measurement made by Bob. The time symmetry inherent in closed systems is not applicable, because of irreversible loss of information into the environment. Our approach has been to find a quite general master equation for the retrodictive density operator, enabling us to determine this operator at the time of state preparation. Projection onto the appropriate preparation device operator then allows us to find the probability that a particular state was prepared by Alice. We emphasise again that the retrodictive formalism is entirely consistent with the more usual predictive formalism of quantum mechanics combined with inference based on Bayes' theorem, but the retrodictive master equation in this paper provides a more direct and natural approach to the quantum communication problem.

We thank the U.K. Engineering and Physical Sciences Research Council and the Australian Research Council for support.

[1]    Y. Aharonov, P. G. Bergman and J. L. Lebowitz, Phys. Rev. **134**, B1410 (1964); R. H. Penfield, Am. J. Phys. **34**, 422 (1966); Y. Aharonov and D. Z. Albert, Phys. Rev. D **29**, 223 (1984); Y. Aharonov and D. Z. Albert, Phys. Rev. D **29**, 228

(1984); Y. Aharonov and L. Vaidman, J. Phys. A: Math. Gen. **24**, 2315 (1991); D. T. Pegg and S. M. Barnett, Quantum and Semiclass. Opt. **1**, 442 (1999).

[2]   *Quantum Communication, Computing and Measurement*, edited by O. Hirota, A. S. Holevo and C. M. Caves (Plenum, New York, 1997); *Quantum Communication, Computing and Measurement 2,* edited by P. Kumar, G.M. D'Ariano and O. Hirota (Kluwer Academic/Plenum, New York, 2000).

[3]   S. J. D. Phoenix and P. D. Townsend, Contemp. Phys. **36**, 165 (1995) and references therein.

[4]   S. M. Barnett, D. T. Pegg, J. Jeffers, O. Jedrkiewicz and R. Loudon Phys. Rev. A **62**, 022313 (2000).

[5]   A more complete analysis of the communication problem can be found in C. E. Shannon and W. Weaver, *The Mathematical Theory of Communication* (University of Illinois Press, Chicago, 1963).

[6]   G. E. P. Box and G. C. Tiao *Bayesian inference in statistical analysis* (Addison-Wesley, Sydney, 1973), p. 10.

[7]   S. M. Barnett, D. T. Pegg and J. Jeffers, J. Mod. Opt., **47**, 1779 (2000).

[8]   S. M. Barnett, D. T. Pegg, J. Jeffers and O. Jedrkiewicz , J. Phys. B: At. Mol. Opt. Phys. **33**, 3047 (2000).

[9] C. W. Helstrom, 1976 *Quantum Detection and Estimation Theory* (Academic Press, New York,1976)

[10] See for example S. M. Barnett and P. M. Radmore *Methods in Theoretical Quantum Optics* (Oxford University Press, Oxford,1997)

[11] G. Lindblad, Commun. Math. Phys. **48**, 119 (1976); S. Stenholm, in *Quantum Dynamics of Simple Systems*, edited by G.-L. Oppo, S. M. Barnett, E. Riis and M. Wilkinson (SUSSP Publications, Edinburgh, 1996).

[12] This follows directly from Eq. (9) and the fact that the elements $\hat{\Pi}_{j,S}$ form a POM.